# Deep Learning-based Automated Diagnosis of Obstructive Sleep Apnea and Sleep Stage Classification in Children Using Millimeter-wave Radar and Pulse Oximeter

*Wei Wang[#][1], Ruobing Song[#][2], Yunxiao Wu[3], Li Zheng[3], Wenyu Zhang[4], Zhaoxi Chen[4], Gang Li[1], Zhifei Xu[2]*

[1] Department of Electronic Engineering, Tsinghua University, Beijing 100084, China

[2] Department of Respiratory, Beijing Children's Hospital, Capital Medical University, National Center for Children's Health, Beijing 100045, China

[3] Department of Otolaryngology Head and Neck Surgery, Beijing Children's Hospital, Capital Medical University, National Children's Medical Center, Beijing 100045, China

[4] Beijing Qinglei Technology Co. Ltd., Beijing 100089, China

[#] Contributed equally.

**Corresponding author:**

Prof Zhifei Xu

Department of Respiratory, Beijing Children's Hospital, Capital Medical University, National Center for Children's Health, Beijing 100045, China

Email: zhifeixu@aliyun.com

Prof Gang Li

Department of Electronic Engineering, Tsinghua University, Beijing 100084, China

Email: gangli@tsinghua.edu.cn

# Abstract

**Study Objectives:** To evaluate the agreement between the millimeter-wave radar-based device and polysomnography (PSG) in diagnosis of obstructive sleep apnea (OSA) and classification of sleep stage in children.

**Methods:** 281 children, aged 1 to 18 years, who underwent sleep monitoring between September and November 2023 at the Sleep Center of Beijing Children's Hospital, Capital Medical University, were recruited in the study. All enrolled children underwent sleep monitoring by PSG and the millimeter-wave radar-based device, QSA600, simultaneously. QSA600 recordings were automatically analyzed using a deep learning model meanwhile the PSG data was manually scored.

**Results:** The Obstructive Apnea-Hypopnea Index (OAHI) obtained from QSA600 and PSG demonstrates a high level of agreement with an intraclass correlation coefficient of 0.945 (95% *CI*: 0.93 to 0.96). Bland-Altman analysis indicates that the mean difference of OAHI between QSA600 and PSG is -0.10 events/h (95% *CI*: -11.15 to 10.96). The deep learning model evaluated through cross-validation showed good sensitivity (81.8%, 84.3% and 89.7%) and specificity (90.5%, 95.3% and 97.1%) values for diagnosing children with OAHI>1, OAHI>5 and OAHI>10. The area under the receiver operating characteristic curve is 0.923, 0.955 and 0.988, respectively. For sleep stage classification, the model achieved Kappa coefficients of 0.854, 0.781, and 0.734, with corresponding overall accuracies of 95.0%, 84.8%, and 79.7% for Wake-sleep classification, Wake-REM-Light-Deep classification, and Wake-REM-N1-N2-N3 classification, respectively.

**Conclusions:** QSA600 has demonstrated high agreement with PSG in diagnosing OSA and performing sleep staging in children. The device is portable, low-load and suitable for follow-up and long-term pediatric sleep assessment.

**Keywords**：obstructive sleep apnea, children, deep learning, sleep stage classification, millimeter-wave radar, portable sleep monitoring device, polysomnography

# Clinical Trials:

# Statement of Significance:

The prevalence of OSA in children ranges from 1.2% to 5.7% and continues to rise annually. Despite its increasing prevalence, OSA in children remains underdiagnosed and inadequately treated in clinical practice. A key contributing factor is the lack of accessible, cost-effective diagnostic tools that are practical for use in multiple settings. Millimeter-wave radar, with its ability to accurately extract motion information, offers a portable, low-impact, and safe alternative. In our study, we evaluated the agreement between the radar-based device and PSG, the gold standard for sleep monitoring, in the automated diagnosis of OSA and sleep stage classification in children. The results showed high agreement between the radar-based device and PSG, confirming its suitability for sleep monitoring in children.

# Introduction

Obstructive sleep apnea (OSA) is characterized by repeated narrowing or obstruction of the upper airway during sleep, resulting in hypopnea, apnea, frequent arousals, and sleep fragmentation. The prevalence of OSA in children ranges from 1.2% to 5.7% and has been increasing annually [1]. Sleep can be categorized into two stages: non-rapid eye movement (NREM) and rapid eye movement (REM). NREM is further divided into three stages: N1, N2, and N3. N1 and N2 are commonly referred to as light sleep, while N3 is recognized as deep sleep. High-quality sleep is characterized by rhythmic cycles alternating between different sleep stages, which is crucial for maintaining brain function. Long-term repeated sleep respiratory events and disruption of sleep architecture have been linked to a series of complications include hypertension, growth retardation, insulin resistance in children [2-5]. In particular, since children are in a critical period of brain development, repeated exposure to intermittent hypoxemia and sleep fragmentation can lead to memory impairment, attention deficit hyperactivity disorder (ADHD) and even long-term neurocognitive disorders [6,7].

Polysomnography (PSG) is considered the gold standard for diagnosing OSA and classifying sleep stages. However, there are several limitations that hinder its widespread application in clinical and daily settings. Traditional PSG utilizes 6-lead electroencephalogram with electrooculogram and electromyogram to assess sleep quality of participants. The use of multiple leads and the resulting discomfort pose significant challenges for its application in young children. Additionally, participants undergoing PSG are required to spend a whole night in a sleep laboratory or other unfamiliar environments. This may make some participants difficult to fall asleep and impact their sleep quality, leading to monitoring results inconsistent with their daily sleeping patterns. Furthermore, the equipment necessary for PSG is expensive, and interpreting and analyzing the results requires specialized expertise from professional technicians. Due to these factors, PSG is not widely utilized domestically or internationally. Consequently, a significant proportion of children with OSA are unable to receive timely diagnosis, treatment or regular follow-up care, which is essential for achieving optimal treatment outcomes.

Radar is a sensor that can extract the characteristics of targets by transmitting and receiving electromagnetic waves. It can accurately perceive human body movement, respiration, and other vital signs without physical contact. This can help participants eliminate the constraints of sensors and cables typically associated with PSG, providing a more comfortable experience. Recent studies have employed radar technology to detect sleep respiratory events and classify sleep stages in adults [8-11], but there are few studies on children with sufficient sample size. Photoplethysmography (PPG) technology enables non-invasive acquisition of physiological data such as heart rate and blood oxygen levels. Numerous studies have leveraged PPG signals to achieve sleep assessment, including automated sleep staging [12] and pediatric OSA screening [13]. PPG provides complementary information to radar data for sleep monitoring, suggesting that combining data from both sensors can improve the performance of sleep assessment.

Our study is a prospective study aimed at developing a deep learning model for the automated sleep assessment in children based on radar and PPG signals. In our study, the model first detects the sleep respiratory events and performs sleep staging according to the signals obtained by the millimeter-wave radar-based sleep monitoring device. Then, the Obstructive Apnea-Hypopnea Index (OAHI) can be calculated for the diagnosis of OSA based on the output of the model. Finally, PSG is used as gold standard method for comparison to assess the performance of the model and evaluate the feasibility of the radar-based device in clinical situations.

## Methods:

### Participants

Children aged 1 year to 18 years who underwent sleep monitoring at the Sleep Center of Beijing Children's Hospital, affiliated with Capital Medical University, between September 2023 and November 2023 were recruited in this study. The exclusion criteria included: the presence of severe cardiovascular and cerebrovascular diseases; severe hepatic, renal or pulmonary insufficiency; long-term or current use of medications such as barbiturates, benzodiazepines, or sedatives; coexisting sleep disorders like insomnia; concurrent CPAP treatment on the trial night; lack of guardian consent for the study; incomplete data. This study was approved by the Medical Ethics Committee of Beijing Children's Hospital (approval number: [2023]-E-092-Y). Children aged 8 years or older and their guardians all signed informed consent. For children under 8 years of age, consent was provided by their guardians on their behalf.

**PSG Monitoring and Analysis of Sleep Respiratory Indexes**

All enrolled children underwent in-lab PSG monitoring (Compumedics E-series, Compumedics, Melbourne, Australia or Alice 5, Respironics, Murrysville, PA, USA or SOMNO screen Plus PSG+ V5 system, SOMNO Medics GmbH, Randersacker, Germany). To ensure the validity of the study, all participants were instructed to refrain from consuming any stimulants such as cola, tea, coffee, sedatives, or hypnotics. Additionally, they were required to sleep for more than 7 hours. The physiological signals from PSG were analyzed and manually scored in accordance with the American Academy of Sleep Medicine (AASM) 2020 scoring criteria [14] by professional technicians, and the results were further confirmed by attending sleep pediatricians.

Apnea is defined as an oronasal flow reduction of greater than 90% for at least two respiratory cycles, accompanied by reduced respiratory efforts throughout the event [14]. Hypopnea is defined as a reduction in airflow of at least ≥30%, accompanied by event-related arousal or oxygen desaturations of higher than or equal to 3%, which persists for a minimum of two respiratory cycles. Obstructive Apnea-hypopnea index (OAHI) is the mean number of obstructive apnea, mixed apnea and obstructive hypopnea events per hour during sleep. According to the Chinese guideline for the diagnosis and treatment of childhood obstructive sleep apnea (2020), the criteria of grading OSA severity in children is shown in Table 1 [15].

**Millimeter-wave Radar-based Sleep Monitoring Device**

All enrolled children who underwent PSG monitoring were simultaneously monitored using the millimeter-wave radar-based device (QSA600, Beijing Qinglei Technology Co., Ltd, Beijing, China). QSA600 is a novel sleep monitoring device that consists of a radar system (BGT60TR13C, Infineon Technologies AG, Neubiberg, Germany) and a fingertip pulse oximeter (MD300W628, Beijing Choice Electronic Technology Co., Ltd, Beijing, China). The monitoring scenario of the radar-based device is shown in Figure 1.

When children experience obstructive sleep apnea or hypopnea events, their respiration significantly weaken, potentially leading to reduced blood oxygen saturation, hypoxia, and decreased heart rate [16]. Furthermore, during different sleep stages, vital signs such as respiratory rate, heart rate, and body movements exhibit distinct patterns [17]. For example, humans typically exhibit a high respiratory rate variability when awake, accompanied by the highest body movement rate. In contrast, during deep sleep, their respiratory rate becomes more stable, while body movement rate decreases significantly. Therefore, by analyzing the respiratory intensity and variability, movement rate, heart rate and other vital signs, the radar-based device could detect sleep respiratory events and classify different sleep stages.

The millimeter-wave radar utilized in our experiment can offer centimeter-level range resolution. Therefore, it enables the differentiation of radar echo signals from various parts of the human body. The received signals are subsequently processed to enable real-time tracking of chest and other body movements. Based on this, the radar system can accurately extract vital signs such as respiratory rate and body movements during sleep.

The pulse oximeter, which can obtain photoplethysmography (PPG) signals, is used to simultaneously monitor children's pulse rate and blood oxygen level during the experiment. The PPG signal measures the variations in blood volume corresponding to each cardiac cycle, which provides different physiological information for sleep assessment compared to the radar signal.

In our algorithm, radar signals are initially preprocessed to generate spectrograms representing different physical significance, including body movement power, breathing effort and breathing doppler. Simultaneously, PPG signals undergo preprocessing to extract features from the time, frequency and time-frequency domain. The deep learning model takes all the above features and spectrograms as the input. It is mainly based on convolutional neural network (CNN) and bidirectional long short-term memory (LSTM) network. CNN can perform further feature extraction on the input, and LSTM enables the model to integrate both past and future information when making predictions at each time step. The model outputs sleep staging results and identifies the temporal locations of detected sleep respiratory events.

**Statistical Analysis**

All statistical analyses were conducted using Python and SPSSPRO. Measurement data were first analyzed for normality using the Shapiro-Wilk test. Normally distributed data were summarized using mean and standard deviation, denoted as $\bar{x} \pm s$. Non-normally distributed data were summarized using median, 25th percentile (P25) and 75th percentile (P75). Categorical data were summarized using count and percentage. The agreement between the two monitoring methods regrading OAHI, total sleep time (TST), and the proportion of each sleep stage was evaluated using the intraclass correlation coefficient (ICC, two-way random-effects model for absolute agreement) and Bland-Altman analysis. Three sleep stage classification tasks were performed and evaluated: 1) Wake-sleep (WS) classification, 2) Wake-REM-Light-Deep (WRLD) classification, and 3) Wake-REM-N1-N2-N3 (WRNN) classification. Frame-by-frame consistency in sleep staging between the two monitoring methods was assessed using metrics of accuracy, Kappa coefficient, recall, and precision. The feasibility of the radar-based device in diagnosing pediatric OSA was quantitatively analyzed using sensitivity and specificity at three different OSA severity thresholds (OAHI >1, >5, and >10 events/h). We applied a 4-fold cross-validation technique splitting the data based on different children, and reported the results on the test set.

## Results

### Participants characteristics

A cohort of 281 children was included in the study, comprising 184 (65%) males and 97 (35%) females. The age distribution is shown in Figure 2. Based on the results of PSG, the enrolled children were categorized into four groups with different OSA severity according to the criteria in Table 1. Demographic and PSG data are presented in Table 2 for reference.

### Agreement of sleep respiratory indexes between the radar-based device and PSG

OAHI is the key indicator in diagnosing OSA in children. In our study, the OAHI values obtained from the radar-based device and PSG demonstrate a high level of agreement (Figure 3A), with an ICC of 0.945 (95% *CI*: 0.93 to 0.96). Bland-Altman analysis (Figure 3B) indicates that the mean difference between radar-based OAHI and PSG OAHI is -0.10 events/h (95% *CI*: -11.15 to 10.96), with 95.0% (267/281) of the data points located within the limits of agreement.

The calculation of OAHI requires sleep monitoring device to distinguish between obstructive and central apnea. Central apnea is characterized by a temporary cessation of respiratory effort due to a malfunction in the respiratory center. Obstructive apnea, on the other hand, involves significant airflow reduction due to airway obstruction, while respiratory effort in the chest and abdomen continues. The radar-based device can detect chest and abdominal movements in children and is able to identify central respiratory events. In our study, the consistency analysis of the Central Apnea Index (CAI) in the radar-based device and PSG showed an ICC of 0.760 (95% *CI*: 0.71 to 0.81), while the mean difference of the Bland-Altman analysis is 0.15 events/h.

**Diagnostic performance of the radar-based device**

Furthermore, we evaluated the performance of the radar-based device in diagnosing OSA in children. Table 3 indicates the diagnostic performance of radar-based device with different OSA severity thresholds. With a threshold of OAHI > 1 events/h, the radar-based device demonstrated a sensitivity of 81.8% (95% *CI*: 76.1 to 87.5%) and a specificity of 90.5% (95% *CI*: 84.9 to 96.1%). When using a threshold of OAHI > 5 events/h, the device exhibited a sensitivity of 84.3% (95% *CI*: 75.8 to 92.8%) and a specificity of 95.3% (95% *CI*: 92.4 to 98.1%). At a higher threshold of OAHI > 10 events/h, the sensitivity increased to 89.7% (95% *CI*: 80.2 to 99.3%) and the specificity to 97.1% (95% *CI*: 95.0 to 99.2%). These results indicate that the radar-based device performs well in OSA diagnosis. Figure 4 presents the receiver operating characteristic (ROC) curves for the radar-based device in OSA diagnosis (OAHI diagnostic thresholds of 1, 5, and 10 events/h), with the areas under the curve (AUC) being 0.923 (95% *CI*: 0.898 to 0.954), 0.955 (95% *CI*: 0.927 to 0.984), and 0.988 (95% *CI*: 0.972 to 1.000), respectively.

**Sleep Staging Performance of the radar-based device**

The confusion matrix for the three different sleep stage classification tasks (WS, WRLD, WRNN) is shown in Figure 5. The results indicate that the radar-based device performs well across all three classification tasks, achieving a recall of 88.9% for wake and 96.7% for sleep. Table 4 presents detailed metrics for the three sleep stage classification tasks. In WS classification, the recall and precision for the wake stage are lower compared to those for the sleep stage. This may be due to the difficulty in detecting brief arousals or insomnia, as well as the tendency to misclassify sleep with movement as wakefulness. In WRLD classification, the recall for the deep sleep stage is slightly lower compared to other categories. This is likely because the physiological characteristics of deep sleep are similar to those of the N2 stage in light sleep, making it easy to misclassification. In WRNN classification, the performance for the N1 stage is poor. The N1 stage typically serves as a transition between wake and N2, resulting in a relatively short duration compared to other stages. This brief duration may hinder the deep learning model's ability to effectively learn the distinctive characteristics of the N1 stage.

We further calculated the TST, light sleep proportion, deep sleep proportion, and REM sleep proportion based on the sleep stage classification results. The TST, light sleep proportion, deep sleep proportion, and REM sleep proportion obtained from the radar-based device and PSG demonstrated good agreement (Figure 6A, 6C, 6E, 6G). The ICC values were 0.906 (95% *CI*: 0.88 to 0.92), 0.581 (95% *CI*: 0.50 to 0.65), 0.621 (95% *CI*: 0.54 to 0.69), and 0.672 (95% *CI*: 0.60 to 0.73), respectively. Bland-Altman analysis (Figure 6B, 6D, 6F, 6H) indicates that the mean differences between radar-based device and PSG measurements for TST, light sleep proportion, deep sleep proportion, and REM sleep proportion were 0.01h (95% *CI*: -0.94 to 0.96), -0.44% (95% *CI*: -14.55 to 13.67), 1.19% (95% *CI*: -10.33 to 12.71), and -0.75% (95% *CI*: -9.72 to 8.22), respectively.

## Discussion

High-quality sleep is essential for the physical and mental health of children. OSA is the most common type of sleep breathing disorder in children, significantly impacting their sleep health and overall quality of life. According to the findings of the seventh national population census, about 13 million children in China may be affected by OSA [18]. If these children cannot get timely diagnosis and treatment, their growth and development could be affected as the condition progresses.

PSG is considered the gold standard technique for diagnosing OSA and classifying sleep stages. However, PSG has limitations in follow-up and long-term monitoring of patients due to its requirement for expensive devices and professional technicians. The application of PSG in young children especially presents a significant challenge in clinical settings. Traditionally, a substantial number of sensors are attached to participants during PSG monitoring. Long-term sensor attachment and frequent sensor detachment can cause considerable discomfort, which in turn affects the accuracy of the monitoring results particularly in young children. Additionally, children's skin is more delicate, and direct contact with sensors could potentially lead to serious damage. As a result, numerous studies are being conducted on developing new sleep monitoring devices, which can promote the prevalence of pediatric sleep assessment.

Millimeter-wave radar has the capability to accurately extract motion information. It can detect millimeter-level chest movements, allowing for precise calculation of important physiological indicators, including body movement rate, respiratory rate, depth of respiration, and other related indicators. Previous studies have utilized radar in the application of sleep assessment, whose results demonstrate a high level of agreement with those obtained from PSG. Zhao et al. evaluated the agreement between in-laboratory PSG and a contact-free radar device in screening for OSA. Using different diagnostic criterion (AHI $\geqslant$ 5, 15, 30 events/h), the device all showed good sensitivity (90.4%, 87.1% and 88.7%) and specificity (77.6%, 89.7% and 95.7%) [8]. Choi et al. placed a 60GHz FMCW radar on the ceiling directly above the bed for sleep monitoring. They designed a model with a convolutional recurrent neural network to detect sleep respiratory events based on the respiratory signals demodulated from the radar signals. Experiment results demonstrated that the estimated AHI showed good agreement (ICC=0.929) with the ground truth [9]. They further developed a hybrid CNN-Transformer architecture and employed dual-source radar technology to enhance the accuracy and performance of OSA diagnosis [10]. de Goederen et al. explored the use of radar for sleep stage classification in children. Using 38 features derived from body movements and breathing rates, an adaptive boosting classifier achieved an accuracy of 89.8% for distinguishing wake and sleep stage [19]. As the sleep stage classification became more detailed, the performance gradually declined, which is not sufficient for clinical application.

The sleep and respiratory characteristics of children differ significantly from those of adults. Compared to adults, children's respiration is characterized by being shallower and faster, with smaller chest and abdominal movements. The diagnostic criteria for pediatric OSA are also different from those for adults. Additionally, parents are often present during sleep monitoring, which can interfere with radar signals. Consequently, diagnosing OSA and performing sleep staging in children is more challenging than in adults. To the best of our knowledge, this study represents the first application of radar technology in diagnosing pediatric OSA and classifying sleep stages using a large sample size.

In this study, we enrolled 281 children to simultaneously undergo monitoring of PSG and the radar-based device. Notably, 16 of these children were between the ages of 1 and 3 years, with the youngest being 1 year and 1 month old. We developed a deep learning model for detecting sleep respiratory events as well as classifying sleep stages in children using millimeter-wave radar-based sleep monitoring device. We trained the model using the collected clinical data and validated its performance by comparing the results with those from PSG. The radar-based device showed high agreement with PSG in OAHI estimation (ICC=0.945). It also demonstrated good performance in diagnosing pediatric OSA across different severity thresholds (OAHI > 1, 5, 10 events/h), achieving sensitivities of 81.8%, 84.3% and 89.7%, and specificities of 90.5%, 95.3% and 97.1%, respectively. Our study also achieved better performance in sleep staging, especially detailed sleep stage classification, than previous studies by expanding the enrollment of participants and optimizing the model. Compared with PSG, our radar-based device showed high accuracy (84.8% and 79.7%) and Cohen's Kappa (0.781 and 0.734) even in WRLD and WRNN classification, which are relatively difficult tasks in sleep staging. These findings suggest that the radar-based device has the potential to be a valuable tool for the sleep assessment in children.

The radar-based device used in our study has the advantages of portability, safety and low load. When undergoing the monitoring of the radar-based device, participants only need to wear pulse oximeter, with the radar system not making direct contact with them. This characteristic greatly enhances the comfort of participants. Furthermore, the radiation energy of the radar system is much lower than that of common electronic consumer products like smartphones and Wi-Fi, which can be safely applied to children.

In conclusion, the radar-based device is suitable for the sleep monitoring needs of children, particularly those in younger age group. We will apply the radar-based device in home settings to evaluate its long-term performance in daily life. Additionally, the radar's capacity to monitor chest movements makes it possible to differentiate between central and obstructive respiratory events. Future research will focus on improving this differentiation capability. We also intend to extend the use of the radar-based device to monitor special pediatric populations, such as children with craniofacial abnormality or neuromuscular diseases. This may expand the application scenario of the radar-based device in the future.

## Limitations

This study demonstrates that the radar-based device shows a high level of agreement with PSG in the diagnosis of OSA and sleep staging. However, it still has several limitations. Firstly, although this study significantly expands the dataset compared to previous research utilizing radar technology for diagnosing OSA in children, the sample size remains limited from a clinical practice perspective. Promoting the clinical application of the radar-based device and extending its usage scenarios will help collect more clinical data, thereby enhancing the performance of the radar-based device. Secondly, the radar-based device exhibits poor performance in distinguishing between N1 and N2 stage. This is because the radar-based device utilizes the differences in respiration, heartbeat, and body movements across various sleep stages to perform sleep staging, while such differences are not apparent between N1 and N2 stage [20]. In contrast, PSG incorporates EEG signals, which provide more precise distinctions. By optimizing the algorithm and model with larger datasets, the radar-based device is expected to improve its ability to distinguish between N1 and N2 stage.

## Conclusion

PSG represents the gold standard for the diagnosis of OSA and the classification of sleep stages. However, multiple shortages limit its large-scale application in clinical settings. The radar-based device has demonstrated good performance in the diagnosis of OSA and the classification of sleep stages in children. This is achieved by analyzing the signals from radar and wrist oximetry and by using models that have been specifically designed for children. In addition, the radar-based device is portable, low-impact and suitable for sleep monitoring in children, particularly young children.

## Funding

This work was supported by National Natural Science Foundation of China under Grant 82070092 and National Natural Science Foundation of China under Grant 61925106.

## Disclosure Statement

The authors report no conflicts of interest in this work.

## Acknowledgments

The authors appreciate the generous work of the linguistic modification by Zhuokang Huang and Peisi Lin.

## Data and source code availability

The data that support the findings of this study are available from the corresponding author upon reasonable request.

# Figure Captions

Figure 1. Monitoring scenario of the radar-based device.

Figure 2. Sex and age distribution of the participants.

Figure 3. Comparison of OAHI between radar-based device and PSG: (A) Scatterplot of OAHI on PSG compared to radar-based device. (B) Bland-Altman Plot of OAHI on PSG compared to radar-based device.

Figure 4. Receiver operating characteristic curves for the radar-based device estimated OAHI vs the PSG OAHI. The 3 curves refer to a diagnostic threshold of the expert annotated OAHI > 1 events/h, OAHI > 5 events/h, and OAHI > 10 events/h. OAHI: obstructive apnea-hypopnea index; PSG: Polysomnography; AUC: area under the curve.

Figure 5. Confusion matrix for sleep stage classification using radar-based device: (A) WS classification results (B) WRLD classification results (C) WRNN classification results.

Figure 6. Comparison of sleep staging results of radar-based device and PSG: (A) Scatterplot of TST (B) Bland-Altman plot of TST (C) Scatterplot of light sleep proportion (D) Bland-Altman plot of light sleep proportion (E) Scatterplot of deep sleep proportion (F) Bland-Altman plot of deep sleep proportion (G) Scatterplot of REM sleep proportion (H) Bland-Altman plot of REM sleep proportion.

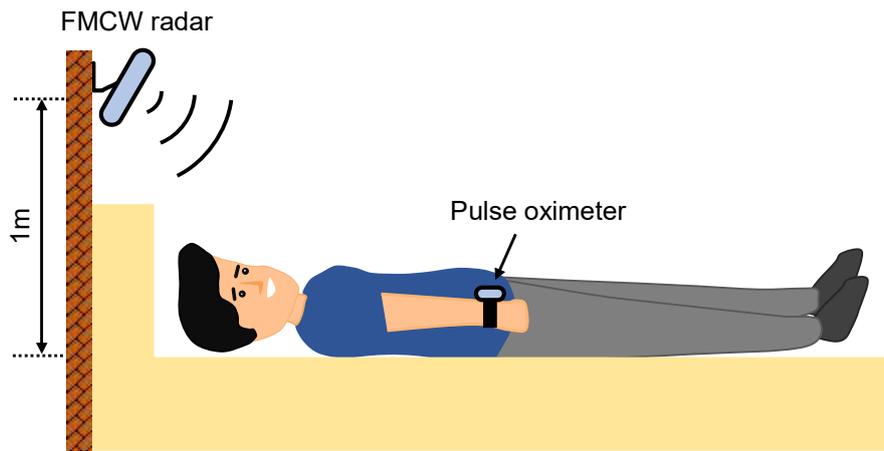
Figure 1

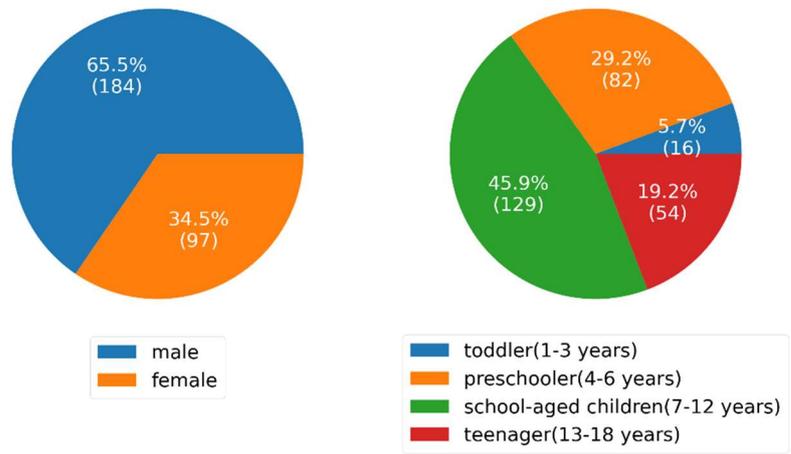

Figure 2

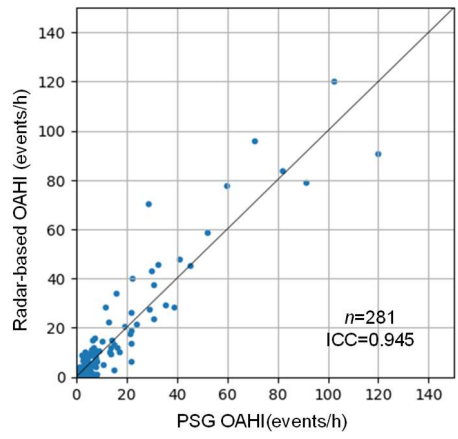 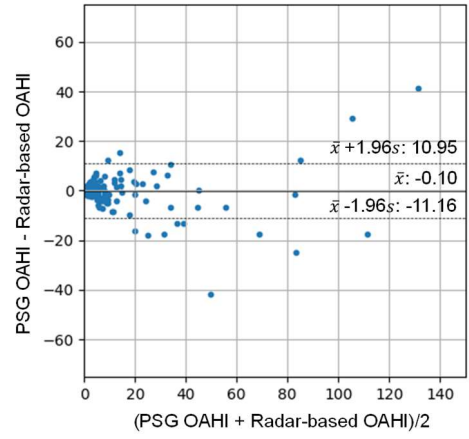

A　　　　　　　　　　　　　　B

Figure 3

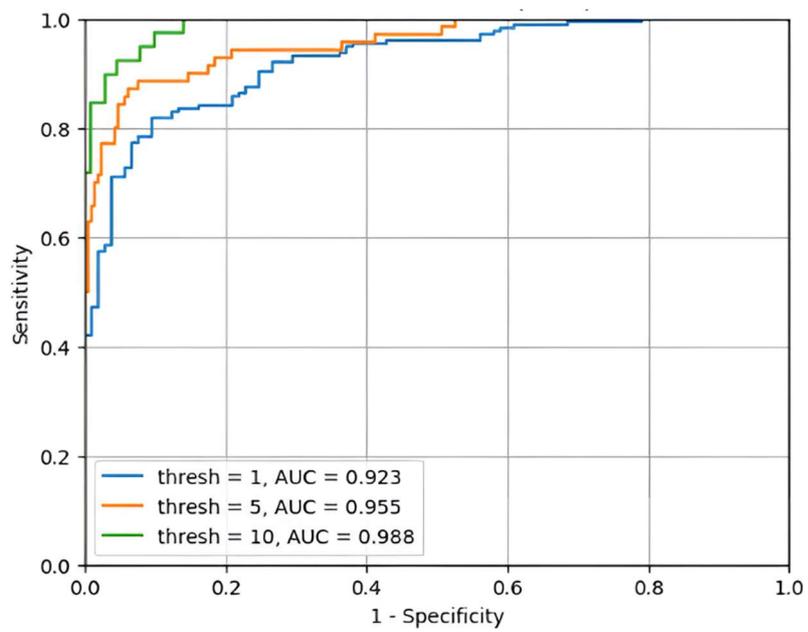

Figure 4

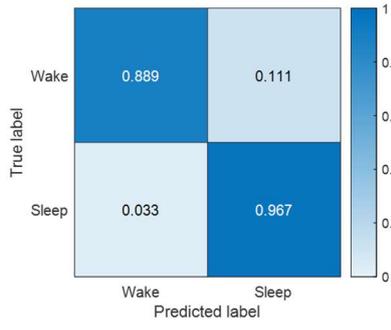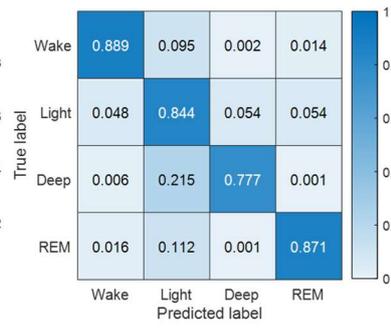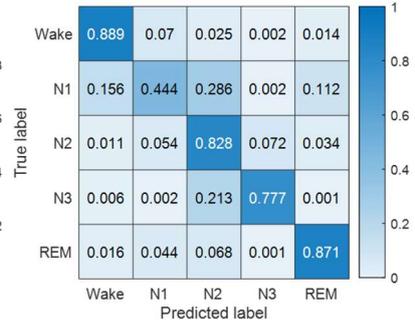

|   A   |   B   |   C   |

Figure 5

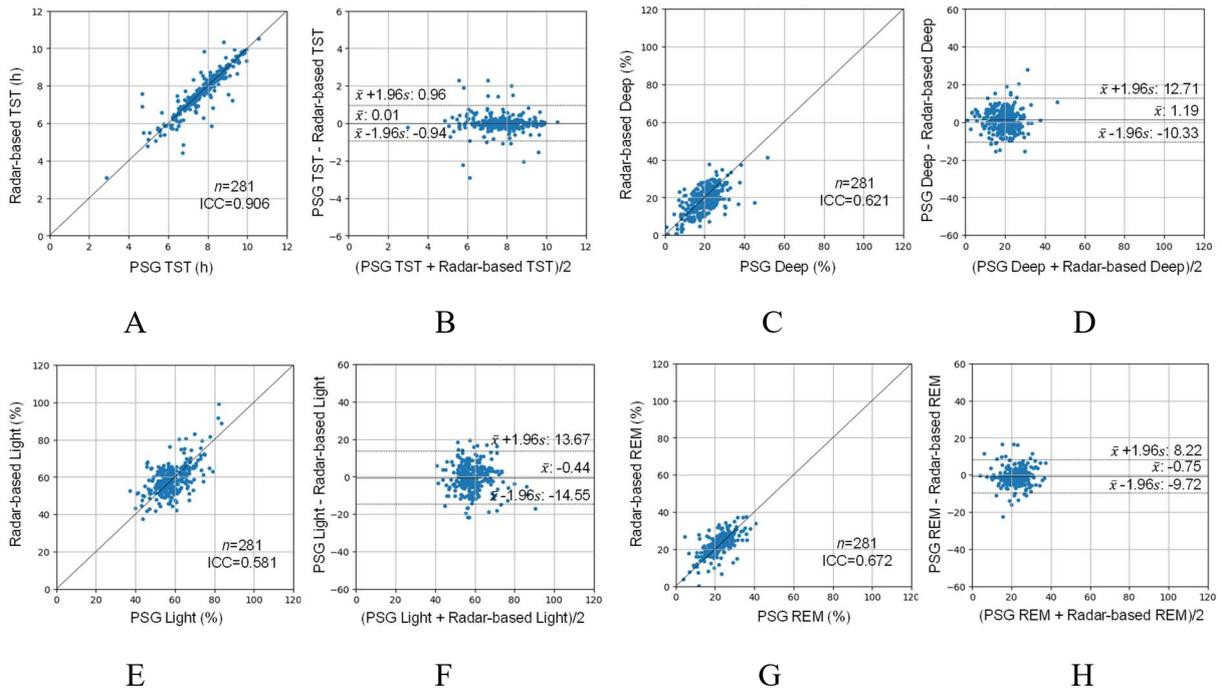

Figure 6

Table 1: The criteria of grading OSA severity in children

| Severity | Healthy | Mild | Moderate | Severe |
|---|---|---|---|---|
| OAHI range (events/h) | OAHI≤1 | 1<OAHI≤5 | 5<OAHI≤10 | OAHI>10 |

OAHI: obstructive sleep apnea-hypopnea index.

Table 2: the demographic and PSG data of enrolled children

| | | Normal | Mild OSA | Moderate OSA | Severe OSA |
|---|---|---|---|---|---|
| Subject characteristic | Number | 105 | 106 | 31 | 39 |
| | Age(year) | 7.25(5.08,9.71) | 8.38(4.73,11.79) | 7.25(4.75,10.83) | 9.58(5.75,12.50) |
| | Gender(male/female) | 71/34 | 62/44 | 21/10 | 29/10 |
| | BMI(kg/m2) | 16.60(14.55,20.45) | 17.60(14.55,25.15) | 18.60(15.30,22.70) | 22.70(18.60,28.60) |
| | OAHI(events/h) | 0.40(0.10,0.60) | 2.37(1.50,3.22) | 6.60(5.60,7.40) | 22.40(15.11,45.30) |
| PSG data | TIB(min) | 542.00(501.70,591.00) | 545.00(491.55,575.75) | 533.00(519.00,579.50) | 535.00(490.30,583.50) |
| | TST(min) | 472.50(442.25,512.75) | 469.50(432.75,514.80) | 461.50(427.50,500.00) | 428.30(372.50,484.00) |
| | Sleep efficiency(%) | 88.90(81.90,93.70) | 89.60(83.70,93.45) | 86.90(74.60,92.50) | 79.60(73.70,92.90) |
| | Sleep latency(min) | 9.50(4.50,10.60) | 9.00(4.50,10.85) | 9.00(5.50,10.50) | 8.50(5.20,11.50) |
| | N1(%) | 10.90(8.10,14.85) | 13.30(8.15,17.00) | 15.70(13.80,17.60) | 19.70(15.00,32.10) |
| | N2(%) | 43.30(39.90,47.85) | 44.60(38.85,51.00) | 41.90(36.60,50.10) | 42.40(34.90,47.80) |
| | N3(%) | 20.80(17.55,24.30) | 19.90(14.40,23.45) | 19.50(14.70,24.40) | 18.30(10.90,24.00) |
| | REM(%) | 22.80(20.05,26.70) | 22.70(19.35,25.10) | 20.80(18.70,24.60) | 16.60(13.80,21.60) |
| | ODI(events/h) | 0.3(0.05,0.80) | 1.20(0.40,2.30) | 3.60(1.60,6.60) | 19.60(10.30,46.60) |
| | Mean O2 saturation(%) | 98.00(98.00,98.00) | 98.00(97.00,98.00) | 98.00(97.00,98.00) | 97.00(95.00,97.00) |
| | Lowest O2 saturation(%) | 94.00(92.00,95.00) | 92.00(90.00,94.00) | 91.00(88.00,93.00) | 85.00(79.00,88.00) |

Data are presented as median(P25, P75). OSA: obstructive sleep apnea; BMI: body mass index; OAHI: obstructive sleep apnea-hypopnea index; TIB: time in bed; TST: total sleep time; N1, N2, N3, and R all referred to as sleep stages. ODI: oxygen desaturation index.

Table 3: Diagnostic performance of the radar-based device with different cutoffs

| OAHI (events/h) | Sensitivity (%) | 95% CI LB(%) | 95% CI UB(%) | Specificity (%) | 95% CI LB(%) | 95% CI UB(%) |
|---|---|---|---|---|---|---|
| > 1 | 81.8 | 76.1 | 87.5 | 90.5 | 84.9 | 96.1 |
| > 5 | 84.3 | 75.8 | 92.8 | 95.3 | 92.4 | 98.1 |
| > 10 | 89.7 | 80.2 | 99.3 | 97.1 | 95.0 | 99.2 |

OAHI: obstructive apnea-hypopnea index; CI: confidence interval; LB: lower bound; UB = upper bound.

Table 4: Detailed metrics for sleep stage classification

| Task | | WS | WRLD | WRNN |
|---|---|---|---|---|
| Recall (%) | Wake | 88.9 | | |
| | N1 | 96.7 | 84.4 | 44.4 |
| | N2 | | | 82.8 |
| | N3 | | 77.7 | |
| | REM | | 87.1 | |
| | Mean | 92.8 | 84.5 | 76.2 |
| Precision (%) | Wake | 88.5 | | |
| | N1 | 96.8 | 84.0 | 55.3 |
| | N2 | | | 77.2 |
| | N3 | | 82.4 | |
| | REM | | 84.4 | |
| | Mean | 92.6 | 84.8 | 77.5 |
| Accuracy (%) | | 95.0 | 84.8 | 79.7 |
| Kappa | | 0.854 | 0.781 | 0.734 |

WS: Wake-sleep classification, WRLD: Wake-REM-Light-Deep classification, WRNN: Wake-REM-N1-N2-N3 classification.